\documentclass[12pt]{article}

\begin{document}

\title{Pfaffian Expressions for Random Matrix 
Correlation Functions}

\author{Taro Nagao}

\date{}
\maketitle

\begin{center}
\it
Graduate School of Mathematics, Nagoya University, 
Chikusa-ku, \\ Nagoya 464-8602, Japan \\
\end{center}

\begin{abstract}
It is well known that Pfaffian formulas for eigenvalue 
correlations are useful in the analysis of real and 
quaternion random matrices. Moreover the parametric 
correlations in the crossover to complex random matrices 
are evaluated in the forms of Pfaffians. In this article, 
we review the formulations and applications of Pfaffian 
formulas. For that purpose, we first present the general 
Pfaffian expressions in terms of the corresponding 
skew orthogonal polynomials. Then we clarify the relation 
to Eynard and Mehta's determinant formula for hermitian 
matrix models and explain how the evaluation is simplified 
in the cases related to the classical orthogonal polynomials. 
Applications of Pfaffian formulas to random matrix theory 
and other fields are also mentioned. 
\end{abstract}

\newpage

\section{Introduction}
\setcounter{equation}{0}
\renewcommand{\theequation}{1.\arabic{equation}}

Pfaffian formulas were first introduced by Dyson to 
the theory of random matrices\cite{DYSON70}. Extending Mehta and 
Gaudin's pioneering work\cite{MG}, Dyson proved that  
the eigenvalue correlation functions for unitary 
random matrices were written in the forms of quaternion 
determinants, which were equivalent to Pfaffians. 
His method was then extended and applied to more 
general random matrix ensembles\cite{MEHTA71,DYSON72,
MEHTA76,MEHTA89,BN,NW}. Among all, Pandey and Mehta's 
work\cite{PM,MP,MEHTA04} on the crossover between matrix 
symmetries revealed the wide applicability of 
Pfaffians. Extending Pandey and Mehta's 
result, Nagao and Forrester established more 
general Pfaffian formulas for parametric 
correlation functions\cite{NF99}. Parametric correlation 
functions describe the correlations among 
the eigenvalues with different crossover parameters. 
Then Pfaffian formulas were furthermore extended 
to involve the multi-matrix models combining 
matrices of different symmetries\cite{NAGAO01}. In 
constructing Pfaffian formulas, the corresponding skew 
orthogonal polynomials play a crucial role. As a result, 
the skew orthogonal polynomials and related ordinary 
orthogonal polynomials are extensively used in the asymptotic 
analysis of the correlations.
\par
Recently Pfaffian formulas found new applications in 
combinatorics and related problems in non-equilibrium 
statistical physics. Vicious random walkers\cite{NF02,NKT,NAGAO03}, 
polynuclear growth model\cite{IS}, sequences of partitions (
Pfaffian Schur process)\cite{BR} and random involutions\cite{FNR} were 
explored and statistical fluctuations were 
analytically evaluated. These new applications 
are regarded as discretizations of random matrix 
ensembles and we often make use of the discrete 
versions of orthogonal polynomials. 
\par
In this article, we present the Pfaffian formulas employed 
in the analysis of eigenvalue correlations, list the 
important special cases and explain the relevance to random 
matrix theory and other applications. In \S 2, the 
general Pfaffian expressions are given in the forms of 
multiple integrals. In \S 3, we demonstrate that in a special 
case the Pfaffians are reduced to a simpler determinant expression. 
In \S 4, we explore simple cases in which the Pfaffian 
formulas become more tractable. In particular, we focus on the cases related 
to the (continuous and discrete) classical orthogonal polynomials and explicitly 
list the corresponding skew orthogonal polynomials. In \S 5, 
the relevance to random matrices and other new applications 
are briefly summarized. 
\par 
To begin with, let us introduce the Pfaffian. For an 
antisymmetric $N \times N$ matrix $A$, when $N$ is even, 
the Pfaffian is defined as
\begin{equation}
{\rm Pf}[A] = \frac{1}{(N/2)!} {\sum_P}' 
(-1)^P A_{j_1 \ j_2} A_{j_3 \ j_4} \cdots A_{j_{N-1} \ j_N}.
\end{equation}
Here $P = (j_1,j_2,\cdots,j_N)$ denotes a permutation 
of $(1,2,\cdots,N)$ and $(-1)^P$ is the sign of $P$. 
The summation ${\sum_P}'$ is taken over all 
$P$ satisfying the restriction $j_1 < j_2$, $j_3 
< j_4$, $\cdots$, $j_{N-1} < j_N$.  
\par  
In terms of the vectors 
\begin{equation}
{\bf x}^{(1)},{\bf x}^{(2)},\cdots,{\bf x}^{(M)} 
\end{equation}
with the elements
\begin{equation}
{\bf x}^{(j)} = 
(x^{(j)}_1, x^{(j)}_2,\cdots, x^{(j)}_N),
\end{equation}
we define the functions
\begin{eqnarray}
\label{peven}
p({\bf x}^{(1)},{\bf x}^{(2)},\cdots,{\bf x}^{(M)}) 
& = & \prod_{j>l}^N (x^{(M)}_j - x^{(M)}_l) \ 
{\rm Pf}[F(x^{(1)}_j,x^{(1)}_l)]_{j,l=1,2,\cdots,N} 
\nonumber \\ & \times &   
\prod_{m=1}^{M-1} 
\det[g^{(m)}(x^{(m+1)}_j,x^{(m)}_l)]_{j,l = 1,2,\cdots,N} 
\end{eqnarray}
for even $N$ and
\begin{eqnarray}
\label{podd}
p({\bf x}^{(1)},{\bf x}^{(2)},\cdots,{\bf x}^{(M)}) 
& = & \prod_{j>l}^N (x^{(M)}_j - x^{(M)}_l)  
\nonumber \\ & \times &   
{\rm Pf}\left[ \begin{array}{cc} 
[F(x^{(1)}_j,x^{(1)}_l)]_{j,l=1,2,\cdots,N} & 
[f(x^{(1)}_j)]_{j=1,2,\cdots,N} \\ 
- [f(x^{(1)}_l)]_{l=1,2,\cdots,N} & 0 \end{array} \right]
\nonumber \\ & \times &   
\prod_{m=1}^{M-1} 
\det[g^{(m)}(x^{(m+1)}_j,x^{(m)}_l)]_{j,l = 1,2,\cdots,N} 
\end{eqnarray}
for odd $N$. We suppose that the corresponding measure 
is given by
\begin{equation}
\prod_{j=1}^N {\rm d}\mu_1(x^{(1)}_j) 
\prod_{j=1}^N {\rm d}\mu_2(x^{(2)}_j) \cdots
\prod_{j=1}^N {\rm d}\mu_M(x^{(M)}_j).
\end{equation}
The functions $F(x,y)$, $g^{(m)}(x,y)$ and $f(x)$ are 
arbitrary besides the antisymmetry relation
\begin{equation}
F(x,y) = - F(y,x)
\end{equation}
and the assumption that divergences are avoided in relevant 
calculations. As explained in \S 5, the functions $p$ (with 
normalization constants multiplied) generalize the probability 
distribution functions of random matrix eigenvalues. 
\par
We are interested in the multiple integrals
\begin{eqnarray}
\label{multiple}
& & W(x^{(1)}_1,\cdots,x^{(1)}_{k_1};x^{(2)}_1,\cdots,x^{(2)}_{k_2}; 
\cdots ;x^{(M)}_1,\cdots,x^{(M)}_{k_M}) = \frac{1}{\prod_{m=1}^M (N - k_m)!} 
\nonumber \\ & \times & \int \prod_{j=k_1 + 1}^N {\rm d}\mu_1(x^{(1)}_j) 
\int \prod_{j=k_2 + 1}^N {\rm d}\mu_2(x^{(2)}_j) \cdots
\int \prod_{j=k_M + 1}^N {\rm d}\mu_M(x^{(M)}_j) \nonumber \\ 
& \times & p({\bf x}^{(1)},{\bf x}^{(2)},\cdots,{\bf x}^{(M)}) ,
\end{eqnarray}
which give the expressions for the eigenvalue correlations. The Pfaffian 
formulas for these multiple integrals are presented in next section. 

\section{Pfaffian Formulas for the Correlation Functions}
\setcounter{equation}{0}
\renewcommand{\theequation}{2.\arabic{equation}}

\subsection{The case $N$ even}

Let us recursively define 
\begin{equation}
\label{gmn}
G^{(m,n)}(x,y) \nonumber \\ 
= \left\{ 
\begin{array}{ll} \delta_m(x-y), & m = n, \\ 
\displaystyle 
\int {\rm d}\mu_{m-1}(z) g^{(m-1)}(x,z) G^{(m-1,n)}(z,y), & m > n, 
\end{array} \right. 
\end{equation}
where $\delta_m(x)$ is Dirac's delta function with respect 
to the measure ${\rm d}\mu_m(x)$. Then we introduce 
\begin{equation}
\label{fmn}
F^{(m,n)}(x,y) = \int {\rm d}\mu_1(x') \int {\rm d}\mu_1(y') G^{(m,1)}(x,x') G^{(n,1)}(y,y') F(x',y')
\end{equation}
and an antisymmetric inner product 
\begin{equation}
\langle  f,g \rangle^{(m)} = \int {\rm d}\mu_m(x) \int {\rm d}\mu_m(y) 
F^{(m,m)}(x,y) f(x) g(y).
\end{equation}
By means of the antisymmetric inner product, we construct the monic skew 
orthogonal polynomials $R^{(M)}_k(x) = x^k + \cdots$, which satisfy the skew 
orthogonality relation 
\begin{eqnarray}
& & \langle R^{(M)}_{2 j},R^{(M)}_{2 l + 1} \rangle^{(M)} = 
- \langle R^{(M)}_{2 l + 1},R^{(M)}_{2 j} \rangle^{(M)} =  r_j \delta_{jl},  \\  
& & \langle R^{(M)}_{2 j},R^{(M)}_{2 l} \rangle^{(M)} = 0, \ \ \ 
\langle R^{(M)}_{2 j+1},R^{(M)}_{2 l + 1} \rangle^{(M)} = 0.
\end{eqnarray}
Let us denote the monic monomial of order $j$ by $\Pi_j(x) = x^j$. Then, 
with a notation
\begin{equation}
\label{jjl}
J_{jl} = \langle \Pi_j, \Pi_l \rangle^{(M)}, 
\end{equation}
we find the determinant expressions for $R^{(M)}_k(x)$ as  
\begin{equation}
\label{rmeven} 
R^{(M)}_{2 k}(x) = \frac{1}{u_k} \left| \begin{array}{ccccc}
\Pi_{2 k}(x) & J_{2 k \ 2 k - 1} & J_{2 k \ 2 k - 2} & \cdots 
& J_{2 k \  0} \\   
\Pi_{2 k - 1}(x) & J_{2 k - 1 \ 2 k - 1} & J_{2 k - 1 \ 2 k - 2} & \cdots 
& J_{2 k - 1 \  0} \\   
\vdots & \vdots & \vdots & \ddots 
& \vdots \\   
\Pi_0(x) & J_{0 \ 2 k - 1} & J_{0 \ 2 k - 2} & \cdots 
& J_{0 \  0} \end{array} \right|,     
\end{equation}
\begin{equation}
\label{rmodd}
R^{(M)}_{2 k + 1}(x) = \frac{1}{u_k} \left| \begin{array}{ccccc}
\Pi_{2 k + 1}(x) & J_{2 k + 1 \ 2 k - 1} & J_{2 k + 1 \ 2 k - 2} & \cdots 
& J_{2 k + 1 \  0} \\   
\Pi_{2 k - 1}(x) & J_{2 k - 1 \ 2 k - 1} & J_{2 k - 1 \ 2 k - 2} & \cdots 
& J_{2 k - 1 \  0} \\   
\vdots & \vdots & \vdots & \ddots 
& \vdots \\   
\Pi_0(x) & J_{0 \ 2 k - 1} & J_{0 \ 2 k - 2} & \cdots 
& J_{0 \  0} \end{array} \right| + v_k R^{(M)}_{2 k}(x),     
\end{equation}
where
\begin{equation}
u_k = \left| \begin{array}{cccc}
J_{2 k - 1 \ 2 k - 1} & J_{2 k - 1 \ 2 k - 2} & \cdots 
& J_{2 k - 1 \  0} \\   
J_{2 k - 2 \ 2 k - 1} & J_{2 k - 2 \ 2 k - 2} & \cdots 
& J_{2 k - 2 \  0} \\   
\vdots & \vdots & \ddots & \vdots \\   
J_{0 \ 2 k - 1} & J_{0 \ 2 k - 2} & \cdots & J_{0 \  0} 
\end{array} \right|     
\end{equation}
and $v_k$ is an arbitrary constant. Note the indeterminacy of 
$R^{(M)}_{2 k + 1}(x)$ due to $v_k$.
\par 
Moreover we introduce 
\begin{equation}
\label{rmk}
R^{(m)}_k(x) = \int {\rm d}\mu_M(y) R^{(M)}_k(y) G^{(M,m)}(y,x),
\end{equation}
\begin{equation}
\label{phimk}
\Phi^{(m)}_k(x) = \int{\rm d}\mu_m(y) R^{(m)}_k(y) F^{(m,m)}(y,x)
\end{equation}
and define the matrices $D^{(m,n)}$, $I^{(m,n)}$, $S^{(m,n)}$ as
\begin{equation}
D^{(m,n)}_{j,l} = \sum_{k=0}^{(N/2)-1} \frac{1}{r_k} \left[  
R^{(m)}_{2 k}(x^{(m)}_j) R^{(n)}_{2 k+1}(x^{(n)}_l) - 
R^{(m)}_{2 k + 1}(x^{(m)}_j) R^{(n)}_{2 k}(x^{(n)}_l) \right],
\end{equation}
\begin{equation}
\label{imn}
I^{(m,n)}_{j,l} = - \sum_{k=0}^{(N/2)-1} \frac{1}{r_k} \left[  
\Phi^{(m)}_{2 k}(x^{(m)}_j) \Phi^{(n)}_{2 k+1}(x^{(n)}_l) - 
\Phi^{(m)}_{2 k + 1}(x^{(m)}_j) \Phi^{(n)}_{2 k}(x^{(n)}_l) \right] + F^{(m,n)}_{j,l},
\end{equation}
\begin{equation}
\label{smn}
S^{(m,n)}_{j,l} = \sum_{k=0}^{(N/2)-1} \frac{1}{r_k} \left[  
\Phi^{(m)}_{2 k}(x^{(m)}_j) R^{(n)}_{2 k+1}(x^{(n)}_l) - 
\Phi^{(m)}_{2 k + 1}(x^{(m)}_j) R^{(n)}_{2 k}(x^{(n)}_l) \right] - G^{(m,n)}_{j,l},
\end{equation}
where
\begin{equation}
F^{(m,n)}_{j,l} = F^{(m,n)}(x^{(m)}_j,x^{(n)}_l)
\end{equation}
and
\begin{equation}
G^{(m,n)}_{j,l} = \left\{ \begin{array}{ll} 0, & m \leq n, \\ 
G^{(m,n)}(x^{(m)}_j,x^{(n)}_l), & m > n.
\end{array} \right.
\end{equation}
Let us consider an antisymmetric matrix $A$ which consists of matrix 
blocks $A^{(m,n)}$, $m,n = 1,2,\cdots,M$. Each block $A^{(m,n)}$ is a 
$2 N \times 2 N$ matrix which consists of $2 \times 2$ blocks
\begin{equation}
A^{(m,n)}_{j,l} = \left( \begin{array}{cc} D^{(m,n)}_{j,l} & S^{(n,m)}_{l,j} \\ 
- S^{(m,n)}_{j,l} & - I^{(m,n)}_{j,l} \end{array} \right), \ \ \ 
j,l = 1,2,\cdots,N.
\end{equation}
Then it turns out that the Pfaffian expression for the multiple integral 
(\ref{multiple}) is\cite{NF99,NAGAO01}
\begin{eqnarray}
\label{pfeven}
& & W(x^{(1)}_1,\cdots,x^{(1)}_{k_1};x^{(2)}_1,\cdots,x^{(2)}_{k_2}; \cdots;
x^{(M)}_1,\cdots,x^{(M)}_{k_M}) \nonumber \\ 
& = & \prod_{j=1}^{N/2} r_{j-1} \ 
{\rm Pf}[A^{(m,n)}(k_m,k_n)]_{m,n=1,2,\cdots,M}.
\end{eqnarray}
Here each block $A^{(m,n)}(k_m,k_n)$ is obtained from $A^{(m,n)}$ 
by removing the $2 k_m+1,2 k_m+2,\cdots,2 N$-th rows and $2 k_n+1,2 k_n+2,
\cdots,2 N$-th columns. The proof of this Pfaffian expression 
is quite similar to that in \cite{NAGAO01}, although it is given 
here in a slightly generalized form.

\subsection{The case $N$ odd}

In terms of the functions $G^{(m,n)}(x,y)$ and $R^{(m)}_k(x)$ 
defined in previous subsection, let us employ the notations
\begin{equation}
f^{(m)}(x) = \int {\rm d}\mu_1(x') G^{(m,1)}(x,x')  f(x')
\end{equation}
and
\begin{equation}
s_k = \int{\rm d}\mu_M(y) R^{(M)}_k(y) f^{(M)}(y).
\end{equation}
Moreover we introduce
\begin{equation}
{\bar R}^{(m)}_k(x) = R^{(m)}_k(x) - \frac{s_k}{s_{N-1}} R^{(m)}_{N-1}(x), \ \ \ k = 0,1,2,\cdots,N-2 
\end{equation}
and
\begin{eqnarray}
{\bar \Phi}^{(m)}_k(x) 
& = & \int{\rm d}\mu_m(y) {\bar R}^{(m)}_k(y) F^{(m,m)}(y,x) \nonumber \\ 
& = & \Phi^{(m)}_k(x) - \frac{s_k}{s_{N-1}} \Phi^{(m)}_{N-1}(x), \ \ \ 
k = 0,1,2,\cdots,N-2.  
\end{eqnarray}
Then the matrices ${\bar D}^{(m,n)}$, ${\bar I}^{(m,n)}$, ${\bar S}^{(m,n)}$ 
are defined as
\begin{equation}
{\bar D}^{(m,n)}_{j,l} = \sum_{k=0}^{(N-3)/2} \frac{1}{r_k} \left[  
{\bar R}^{(m)}_{2 k}(x^{(m)}_j) {\bar R}^{(n)}_{2 k+1}(x^{(n)}_l) - 
{\bar R}^{(m)}_{2 k + 1}(x^{(m)}_j) {\bar R}^{(n)}_{2 k}(x^{(n)}_l) \right],
\end{equation}
\begin{eqnarray}
{\bar I}^{(m,n)}_{j,l} & = & - \sum_{k=0}^{(N-3)/2} \frac{1}{r_k} \left[  
{\bar \Phi}^{(m)}_{2 k}(x^{(m)}_j) {\bar \Phi}^{(n)}_{2 k+1}(x^{(n)}_l) - 
{\bar \Phi}^{(m)}_{2 k + 1}(x^{(m)}_j) {\bar \Phi}^{(n)}_{2 k}(x^{(n)}_l) \right]  
\nonumber \\ 
& + &  \frac{1}{s_{N-1}}  \left[ \Phi^{(m)}_{N-1}(x^{(m)}_j) f^{(n)}(x^{(n)}_l) - 
\Phi^{(n)}_{N-1}(x^{(n)}_l) f^{(m)}(x^{(m)}_j) \right] + F^{(m,n)}_{j,l}, 
\nonumber \\ 
\end{eqnarray}
\begin{eqnarray}
{\bar S}^{(m,n)}_{j,l} & = & \sum_{k=0}^{(N-3)/2} \frac{1}{r_k} \left[  
{\bar \Phi}^{(m)}_{2 k}(x^{(m)}_j) {\bar R}^{(n)}_{2 k+1}(x^{(n)}_l) - 
{\bar \Phi}^{(m)}_{2 k + 1}(x^{(m)}_j) {\bar R}^{(n)}_{2 k}(x^{(n)}_l) \right] 
\nonumber \\  
& + & \frac{1}{s_{N-1}} f^{(m)}(x^{(m)}_j) R^{(n)}_{N-1}(x^{(n)}_l) - G^{(m,n)}_{j,l}.
\end{eqnarray}
As before we construct an antisymmetric matrix ${\bar A}$ so that it consists of matrix 
blocks ${\bar A}^{(m,n)}$, $m,n = 1,2,\cdots,M$ and each block consists of  
$2 \times 2$ blocks
\begin{equation}
{\bar A}^{(m,n)}_{j,l} = \left( \begin{array}{cc} {\bar D}^{(m,n)}_{j,l} & {\bar S}^{(n,m)}_{l,j} \\  - {\bar S}^{(m,n)}_{j,l} & - {\bar I}^{(m,n)}_{j,l} \end{array} \right), \ \ \ 
j,l = 1,2,\cdots,N.
\end{equation}
It follows that the Pfaffian expression for the multiple 
integral (\ref{multiple}) is written as\cite{NF99,NAGAO01}
\begin{eqnarray}
& & W(x^{(1)}_1,\cdots,x^{(1)}_{k_1};x^{(2)}_1,\cdots,x^{(2)}_{k_2}; \cdots;
x^{(M)}_1,\cdots,x^{(M)}_{k_M}) \nonumber \\ 
& = & s_{N-1} \ \prod_{j=1}^{(N-1)/2} r_{j-1} \ 
{\rm Pf}[{\bar A}^{(m,n)}(k_m,k_n)]_{m,n=1,2,\cdots,M}.
\end{eqnarray}
One obtains each block ${\bar A}^{(m,n)}(k_m,k_n)$ by removing the $2 k_m+1,2 
k_m+2,\cdots,2 N$-th rows and $2 k_n+1,2 k_n+2,\cdots,2 N$-th columns from 
the block ${\bar A}^{(m,n)}$. It is again straightforward to prove this 
Pfaffian formula by following the procedures in \cite{NAGAO01}
\par
We remark that the above Pfaffian formulas can 
be extended to the case with the factor 
\begin{equation}
\prod_{j>l}^N (x^{(M)}_j - x^{(M)}_l) = 
{\rm det}[(x^{(M)}_j)^{l-1}]_{j,l=1,2,\cdots,N}
\end{equation}
(in (\ref{peven}) and (\ref{podd})) replaced by a 
general determinant factor\cite{BR}
\begin{equation}
{\rm det}[\varphi_{l-1}(x^{(M)}_j)]_{j,l=1,2,\cdots,N}.
\end{equation}
In that case we need to replace $\Pi_j(x)$ by $\varphi_j(x)$ 
in (\ref{jjl}), (\ref{rmeven}) and (\ref{rmodd}). The resulting 
skew orthogonal functions $R_k^{(M)}(x)$ are not in general polynomials. 

\section{Reduction to Eynard-Mehta Formula}
\setcounter{equation}{0}
\renewcommand{\theequation}{3.\arabic{equation}}

In this section we consider the special case
\begin{equation}
\label{gspecial}
g^{(1)}(x,y) = \sum_{j=0}^{N-1} \frac{1}{h_j} Q_j(x) R_j(y),
\end{equation}
where $Q_j(x) = x^j + \cdots$ and $R_j(x) = x^j + 
\cdots$ are monic polynomials.  The determinant of this function 
can be rewritten as
\begin{eqnarray}
& & {\rm det}[g^{(1)}(x_j,y_l)]_{j,l=1,2,\cdots,N} \nonumber \\ & =  & 
\frac{1}{\prod_{j=0}^{N-1} h_j} \det[Q_{j-1}(x_l)]_{j,l = 1,2,\cdots,N} 
\det[R_{j-1}(y_l)] _{j,l=1,2,\cdots,N} \nonumber \\  
& =  & \frac{1}{\prod_{j=0}^{N-1} h_j} \det[(x_l)^{j-1}]_{j,l = 1,2,\cdots,N} 
\det[(y_l)^{j-1}] _{j,l=1,2,\cdots,N} \nonumber \\  
& = &  \frac{1}{\prod_{j=0}^{N-1} h_j}  \prod_{j<l}^{N} (x_j - x_l) 
\prod_{j<l}^N (y_j - y_l) ,
\end{eqnarray}
so that both of the functions (\ref{peven}) and (\ref{podd}) are decomposed 
into two 
factors as
\begin{equation}
p({\bf x}^{(1)},{\bf x}^{(2)},\cdots,{\bf x}^{(M)}) =
p^{({\rm I})} ({\bf x}^{(1)}) \ p^{({\rm II})}({\bf x}^{(2)},\cdots,{\bf x}^{(M)}) .
\end{equation}
The second factor 
\begin{eqnarray}
& & p^{({\rm II})}({\bf x}^{(2)},\cdots,{\bf x}^{(M)}) =
\frac{1}{\prod_{j=0}^{N-1} h_j} \prod_{j>l}^N (x^{(M)}_j - 
x^{(M)}_l) (x^{(2)}_j - x^{(2)}_l)  \nonumber \\ & \times  & \prod_{m=2}^{M-1} {\rm det}[ 
g^{(m)}(x^{(m+1)}_j,x^{(m)}_l)]_{j,l=1,2,\cdots,N}
\end{eqnarray}
has the form of the probability distribution function for the hermitian 
multi-matrix models. Eynard and Mehta derived a determinant formula for 
the eigenvalue correlations of such hermitian multi-matrix models\cite{EM}. 
Therefore, in this special case, their determinant formula should be 
reclaimed from the Pfaffian formulas.
\par
In order to see the reduction to Eynard and Mehta's formula, we choose $Q_j(x)$ 
and constants $h_j$ so that the orthogonality relation
\begin{equation}
\int {\rm d}\mu_M(x) {\rm d}\mu_2(y) G^{(M,2)}(x,y) P_j(x) Q_l(y)  = h_j \delta_{jl}
\end{equation}
holds. Here $P_j(x) = x^j + \cdots$ are monic polynomials. Moreover we define
\begin{equation}
P^{(m)}_j(x) = \int {\rm d}\mu_M(y) P_j(y) G^{(M,m)}(y,x), \ \ \ 1 \leq m \leq M   
\end{equation}
and
\begin{equation}
Q^{(m)}_j(x) = \int{\rm d}\mu_2(y) G^{(m,2)}(x,y) Q_j(y), \ \ \ 2 \leq m \leq M.
\end{equation}
Then it can readily be seen that
\begin{equation}
\int {\rm d}\mu_m(x) \int {\rm d}\mu_n(y) G^{(m,n)}(x,y) P^{(m)}_j(x) Q^{(n)}_l(y)  
= h_j \delta_{jl}.
\end{equation}
\par
Let us use (\ref{gspecial}) to find
\begin{eqnarray}
G^{(m,1)}(x,y) & = & \int {\rm d}\mu_2(z) G^{(m,2)}(x,z) g^{(1)}(z,y) 
\nonumber \\ & = & 
\sum_{j=0}^{N-1} \frac{1}{h_j} Q^{(m)}_j(x) R_j(y), \ \ \ 2 \leq m \leq M. 
\end{eqnarray}
Then we obtain
\begin{eqnarray}
F^{(m,n)}(x,y) & = & \int {\rm d}\mu_1(x') 
\int {\rm d}\mu_1(y') G^{(m,1)}(x,x') G^{(n,1)}(y,y') F(x',y') 
\nonumber \\  & = & \sum_{j=0}^{N-1} \sum_{l=0}^{N-1} 
\frac{1}{h_j h_l} Q^{(m)}_j(x) Q^{(n)}_l(y) 
\langle R_j,R_l \rangle^{(1)}, \ \ \ 2 \leq m,n \leq M, \nonumber \\  
\end{eqnarray}
from which it follows that
\begin{equation}
\label{pprr}
\int {\rm d}\mu_M(x) \int {\rm d}\mu_M(y) F^{(M,M)}(x,y) P^{(M)}_j(x) P^{(M)}_l(y) = 
\langle R_j,R_l \rangle^{(1)}.
\end{equation}
\par
Let us choose $R_k(x)$ so that the skew orthogonality relation
\begin{eqnarray}
& & \langle R_{2 j},R_{2 l + 1} \rangle^{(1)} = 
- \langle R_{2 l + 1},R_{2 j} \rangle^{(1)} =  r_j \delta_{jl},  \\  
& & \langle R_{2 j},R_{2 l} \rangle^{(1)} = 0, \ \ \ 
\langle R_{2 j+1},R_{2 l + 1} \rangle^{(1)} = 0
\end{eqnarray}
holds. Then it can immediately be seen from (\ref{pprr}) that
\begin{equation}
R^{(M)}_j(x) = P^{(M)}_j(x),
\end{equation}
which yields
\begin{equation}
R^{(m)}_j(x) = P^{(m)}_j(x), \ \ \ m = 1,2,\cdots,M
\end{equation}
in general. Moreover we find
\begin{eqnarray}
\label{phipf}
\Phi^{(m)}_j(x) & = & \int {\rm d}\mu_m(y) P^{(m)}_j(y) F^{(m,m)}(y,x) 
\nonumber \\  & = & \sum_{l=0}^{N-1} \frac{1}{h_l} Q^{(m)}_l(x) 
\langle R_j,R_l \rangle^{(1)}, \ \ \ 2 \leq m \leq M,
\end{eqnarray}
which means
\begin{eqnarray}
\label{phi}
\Phi^{(m)}_{2 j}(x) = 
\frac{r_j}{h_{2 j + 1}} Q^{(m)}_{2 j + 1}(x), \ \ \ 2 j \leq N - 2, \ \ \ 
2 \leq m \leq M,   
\nonumber \\ 
\Phi^{(m)}_{2 j + 1}(x) = -  \frac{r_j}{h_{2 j }} Q^{(m)}_{2 j}(x), \ \ \ 
2 j + 1 \leq N, \ \ \ 
2 \leq m \leq M. 
\end{eqnarray}
Therefore, for even $N$, it follows from (\ref{imn}) and (\ref{phi}) that
\begin{equation}
I^{(m,n)}_{j,l} = 0, 
\ \ \ 2 \leq m,n \leq M
\end{equation}
and
\begin{equation}
I^{(m,1)}_{j,l} = - I^{(1,m)}_{l,j} = 0, \ \ \ 2 \leq m \leq M.
\end{equation}
Moreover, from (\ref{smn}) and (\ref{phi}), we obtain
\begin{equation}
\label{smnsp}
S^{(m,n)}_{j,l} = \sum_{k=0}^{N-1} \frac{1}{h_k} Q^{(m)}_k(x^{(m)}_j) 
P^{(n)}_k(x^{(n)}_l) - G^{(m,n)}_{j,l}, \ \ \ 2 \leq m,n \leq M
\end{equation}
and
\begin{equation}
S^{(m,1)}_{j,l} = 0, \ \ \ m \geq 2.
\end{equation}
\par
We are now in a position to examine the Pfaffian expression (\ref{pfeven}). 
It follows from the properties of the Pfaffian that the elements of 
the blocks $A^{(m,n)}(k_m,k_n)$ 
\begin{equation}
A^{(m,n)}_{j,l} = \left\{ \begin{array}{ll} 
\left( \begin{array}{cc} D^{(m,n)}_{j,l} & S^{(n,m)}_{l,j} \\ 
- S^{(m,n)}_{j,l} & 0 \end{array} \right), & m,n \geq 2, \\ 
\left( \begin{array}{cc} D^{(m,n)}_{j,l} & S^{(n,m)}_{l,j} \\ 
0 & 0 \end{array} \right), & m \geq 2,\ n = 1, \\ 
\left( \begin{array}{cc} D^{(m,n)}_{j,l} & 0 \\ 
- S^{(m,n)}_{j,l} & 0 \end{array} \right), & m=1, \ n \geq 2
\end{array} \right.
\end{equation}
can be replaced by
\begin{equation}
A^{(m,n)}_{j,l} = \left\{ \begin{array}{ll} 
\left( \begin{array}{cc} 0 & S^{(n,m)}_{l,j} \\ 
- S^{(m,n)}_{j,l} & 0 \end{array} \right), & m,n \geq 2, \\ 
\left( \begin{array}{cc} 0 & 0 \\ 
0 & 0 \end{array} \right), & m \geq 2,\ n = 1, \\ 
\left( \begin{array}{cc} 0 & 0 \\ 0 
& 0 \end{array} \right), & m=1, \ n \geq 2 
\end{array} \right.
\end{equation}
without changing the value of the Pfaffian. Therefore it is 
straightforward to see that 
\begin{eqnarray}
& & {\rm Pf}[A^{(m,n)}(k_m,k_n)]_{m,n=1,2,\cdots,M} \nonumber \\ 
& = & {\rm Pf}[A^{(1,1)}(k_1,k_1)] \ {\rm Pf}\left[ \begin{array}{cc} 
0 & S^{(n,m)}_{l,j} \\ - S^{(m,n)}_{j,l} & 0 \end{array} 
\right]_{j=1,\cdots,k_m,l=1,\cdots,k_n,m=2,\cdots,M,n=2,\cdots,M} 
\nonumber \\  
& = & {\rm Pf}[A^{(1,1)}(k_1,k_1)] \ 
\det\left[S^{(m,n)}_{j,l}\right]_{j=1,\cdots,k_m,l=1,\cdots,k_n,
m=2,\cdots,M,n=2,\cdots,M}.
\end{eqnarray}
The expression (\ref{smnsp}) of $S^{(m,n)}_{j,l}$ reveals that the 
second determinant factor in the last line of the above equation 
gives Eynard and Mehta's formula.  
\par
For odd $N$, (\ref{phipf}) yields   
\begin{equation}
\Phi^{(m)}_{N-1}(x) = 0, \ \ \ 2 \leq m \leq M, 
\end{equation}
so that
\begin{equation}
{\bar \Phi}^{(m)}_k(x) = \Phi^{(m)}_k(x), \ \ \ 2 \leq m \leq M, 
\end{equation}
from which we can readily find 
\begin{equation}
{\bar I}^{(m,n)}_{j,l} = 0, \ \ \ 2 \leq m,n \leq M,
\end{equation}
\begin{equation}
{\bar I}^{(m,1)}_{j,l} = - {\bar I}^{(1,m)}_{l,j} = 0, \ \ \ 2 \leq m \leq M,
\end{equation}
\begin{equation} 
{\bar S}^{(m,n)}_{j,l} = \sum_{k=0}^{N-1} \frac{1}{h_k} Q^{(m)}_k(x^{(m)}_j) 
P^{(n)}_k(x^{(n)}_l) - G^{(m,n)}_{j,l}, \ \ \ 2 \leq m,n \leq M
\end{equation}
and
\begin{equation}
{\bar S}^{(m,1)}_{j,l} = 0, \ \ \ m \geq 2.
\end{equation}
Following the steps as before, we can readily see that 
\begin{eqnarray}
& & {\rm Pf}[{\bar A}^{(m,n)}(k_m,k_n)]_{m,n=1,2,\cdots,M} \nonumber \\ 
& = & {\rm Pf}[{\bar A}^{(1,1)}(k_1,k_1)] \ 
\det\left[{\bar S}^{(m,n)}_{j,l}\right]_{j=1,\cdots,k_m,l=1,\cdots,k_n,
m=2,\cdots,M,n=2,\cdots,M}, \nonumber \\   
\end{eqnarray}
which gives Eynard and Mehta's formula.
\par
Thus we have shown that Eynard and Mehta's determinant 
formula in \cite{EM} can be derived from the special case 
(\ref{gspecial}) of the Pfaffian formulas in \S 2.    

\section{Simple Special Cases}
\setcounter{equation}{0}
\renewcommand{\theequation}{4.\arabic{equation}}

Let us go back to the general Pfaffian formulas in \S 2. The key ingredients 
of the Pfaffian expressions are the functions $R^{(m)}_j(x)$ 
generated by (\ref{rmk}) 
from the monic skew orthogonal polynomials $R^{(M)}_j(x)$. We introduce the monic 
orthogonal polynomials $C^{(m)}_j(x) = x^j + \cdots$ with the 
orthogonality relation
\begin{equation}
\int {\rm d}\mu_m(x) C^{(m)}_j(x) C^{(m)}_l(x) = h^{(m)}_j \delta_{jl}.
\end{equation}
Now, if the function $g^{(m)}(x,y)$ is expanded as 
\begin{equation}
\label{gex}
g^{(m)}(x,y) = \sum_{j=0}^{\infty} \frac{\gamma^{(m+1)}_j}{\gamma^{(m)}_j} 
\frac{C^{(m+1)}_j(x) C^{(m)}_j(y)}
{\sqrt{h^{(m+1)}_j h^{(m)}_j}}, 
\end{equation}
the evaluations of the integrals in (\ref{gmn}) and (\ref{rmk}) 
are greatly simplified\cite{NF99,NF98}. In this section, 
we assume this simplifying property.
\par
By means of the expansion (\ref{gex}), $G^{(m,n)}(x,y)$ 
is evaluated as 
\begin{equation}
\label{gmnex}
G^{(m,n)}(x,y) = \sum_{j=0}^{\infty} \frac{\gamma^{(m)}_j}{
\gamma^{(n)}_j} \frac{C^{(m)}_j(x) C^{(n)}_j(y)}
{\sqrt{h^{(m)}_j h^{(n)}_j}}, 
\end{equation}
where we assume the completeness of $C^{(m)}_j(x)$ 
\begin{equation}
\delta_m(x-y) = \sum_{j=0}^{\infty} \frac{C^{(m)}_j(x) C^{(m)}_j(y)}
{h^{(m)}_j}. 
\end{equation}
Therefore, if we expand $R^{(M)}_k(x)$ in terms of 
$C^{(M)}_j(x)$ as
\begin{equation}
R^{(M)}_k(x) = \sum_{j=0}^k \alpha_{kj} \frac{C^{(M)}_j(x)}{\gamma^{(M)}_j 
\sqrt{h^{(M)}_j}}, \ \ \ \alpha_{kk} = \gamma^{(M)}_k \sqrt{h^{(M)}_k},
\end{equation}   
it can be readily seen that
\begin{equation}
\label{rmkex}
R^{(m)}_k(x) = \sum_{j=0}^k \alpha_{kj} \frac{C^{(m)}_j(x)}{\gamma^{(m)}_j 
\sqrt{h^{(m)}_j}}, \ \ \ m = 1,2,\cdots,M 
\end{equation}   
in general. Note that $\alpha_{kj}$ are independent of $m$. 
Therefore, if $\alpha_{kj}$ are known, $R^{(m)}_k(x)$ are 
specified for all $m$. It is also possible to write down the 
inverse expansion as 
\begin{equation}
\label{cmkex}
\frac{C^{(m)}_k(x)}{\gamma^{(m)}_k 
\sqrt{h^{(m)}_k}} = \sum_{j=0}^k \beta_{kj} R^{(m)}_j(x). 
\end{equation}
\par
Putting (\ref{gmnex}) into (\ref{fmn}) leads to
\begin{equation}
\label{fmnex}
F^{(m,n)}(x,y) = \sum_{j=0}^{\infty} \sum_{l=0}^{\infty} 
\frac{\gamma^{(m)}_j \gamma^{(n)}_l}{\gamma^{(1)}_j \gamma^{(1)}_l}
\frac{C^{(m)}_j(x) C^{(n)}_l(y)}{
\sqrt{h^{(m)}_j h^{(1)}_j h^{(n)}_l h^{(1)}_l}} 
\langle C^{(1)}_j,C^{(1)}_l \rangle^{(1)}.
\end{equation} 
Substituting (\ref{fmnex}) into (\ref{phimk}) and using 
the expansions (\ref{rmkex}) and (\ref{cmkex}), we obtain
\begin{eqnarray}
\Phi^{(m)}_k(x) & = & \sum_{l=0}^{\infty} \frac{\gamma^{(m)}_l}{\gamma^{(1)}_l} 
\frac{C^{(m)}_l(x)}{\sqrt{h^{(m)}_l h^{(1)}_l}} \langle 
R^{(1)}_k, C^{(1)}_l \rangle^{(1)} \nonumber \\  
& = & \sum_{\nu=0}^{\infty} \sum_{l = \nu}^{\infty} 
\gamma^{(m)}_l \frac{C^{(m)}_l(x)}{\sqrt{h^{(m)}_l}} 
\beta_{l \nu}  \langle 
R^{(1)}_k, R^{(1)}_\nu \rangle^{(1)},
\end{eqnarray}  
from which it follows that   
\begin{eqnarray}
\label{phimkex}
\Phi^{(m)}_{2 k}(x)  
& = & \sum_{l=2 k + 1}^{\infty}  
\gamma^{(m)}_l 
\frac{C^{(m)}_l(x)}{\sqrt{h^{(m)}_l}} 
\beta_{l \ 2 k + 1} r_k, \nonumber \\  
\Phi^{(m)}_{2 k + 1}(x)  
& = & - \sum_{l=2 k}^{\infty}  
\gamma^{(m)}_l 
\frac{C^{(m)}_l(x)}{\sqrt{h^{(m)}_l}} 
\beta_{l \ 2 k} r_k.
\end{eqnarray}
\par
For even $N$, let us put (\ref{phimkex}) into (\ref{smn}). 
After some algebra we arrive at
\begin{eqnarray}
S^{(m,n)}_{j,l} & = & \sum_{k=0}^{N-1} \frac{\gamma^{(m)}_k}{
\gamma^{(n)}_k} \frac{C^{(m)}_k(x^{(m)}_j) C^{(n)}_k(x^{(n)}_l)}
{\sqrt{h^{(m)}_k h^{(n)}_k}} \nonumber \\ 
& + & \sum_{k=0}^{N-1} \sum_{\nu=N}^{\infty} \gamma^{(m)}_{\nu} 
\frac{C^{(m)}_{\nu}(x^{(m)}_j)}
{\sqrt{h^{(m)}_{\nu}}} \beta_{\nu k} R^{(n)}_k(x^{(n)}_l) 
- G^{(m,n)}_{j,l}.
\end{eqnarray}
We also see that the expansion (\ref{cmkex}) can be substituted 
into (\ref{fmnex}). A comparison with (\ref{phimkex}) yields 
\begin{equation}
F^{(m,n)}(x,y) = 
\sum_{k=0}^{\infty} \frac{1}{r_k} \left[ 
\Phi^{(m)}_{2 k}(x) \Phi^{(n)}_{2 k + 1}(y) - 
\Phi^{(m)}_{2 k + 1}(x) \Phi^{(n)}_{2 k}(y) \right], 
\end{equation}
so that 
\begin{equation}
I^{(m,n)}_{j,l} = 
\sum_{k=N/2}^{\infty} \frac{1}{r_k} \left[ 
\Phi^{(m)}_{2 k}(x^{(m)}_j) \Phi^{(n)}_{2 k + 1}(x^{(n)}_l) -   
\Phi^{(m)}_{2 k + 1}(x^{(m)}_j) \Phi^{(n)}_{2 k}(x^{(n)}_l) \right].
\end{equation}
\par
For odd $N$, noting
\begin{equation}
f^{(m)}(x) = \sum_{k=0}^{\infty} s_k \sum_{\nu = k}^{\infty} 
\gamma^{(m)}_{\nu} \frac{C^{(m)}_{\nu}(x)}{\sqrt{h^{(m)}_{\nu}}} 
\beta_{\nu k} = \sum_{k=0}^{\infty} \frac{1}{r_k} \left[ 
\Phi^{(m)}_{2 k}(x) s_{2 k + 1} -   
\Phi^{(m)}_{2 k + 1}(x) s_{2 k} \right],
\end{equation}
we can readily find
\begin{eqnarray}
{\bar S}^{(m,n)}_{j,l} & = & \sum_{k=0}^{N-1} \frac{\gamma^{(m)}_k}{
\gamma^{(n)}_k} \frac{C^{(m)}_k(x^{(m)}_j) C^{(n)}_k(x^{(n)}_l)}
{\sqrt{h^{(m)}_k h^{(n)}_k}} \nonumber \\ 
& + & \sum_{k=0}^{N-1} \sum_{\nu=N}^{\infty} \gamma^{(m)}_{\nu} 
\frac{C^{(m)}_{\nu}(x^{(m)}_j)}
{\sqrt{h^{(m)}_{\nu}}} \beta_{\nu k} R^{(n)}_k(x^{(n)}_l) 
- G^{(m,n)}_{j,l} \nonumber \\ 
& - & \frac{\Phi^{(m)}_{N-1}(x)}{s_{N-1}}
\sum_{k=0}^{(N-3)/2} \frac{1}{r_k} \left[ 
s_{2 k} R^{(n)}_{2 k + 1}(y) -  
s_{2 k + 1} R^{(n)}_{2 k}(y) \right] \nonumber \\ 
& + & \frac{R^{(n)}_{N-1}(y)}{s_{N-1}} 
\sum_{k=N}^{\infty} s_k \sum_{\nu = k}^{\infty} 
\gamma^{(m)}_{\nu} \frac{C^{(m)}_{\nu}(x)}{\sqrt{h^{(m)}_{\nu}}} 
\beta_{\nu k}
\end{eqnarray}  
and 
\begin{eqnarray}
{\bar I}^{(m,n)}_{j,l} & = &  
\sum_{k=(N+1)/2}^{\infty} \frac{1}{r_k} \left[ 
\Phi^{(m)}_{2 k}(x^{(m)}_j) \Phi^{(n)}_{2 k + 1}(x^{(n)}_l) -   
\Phi^{(m)}_{2 k + 1}(x^{(m)}_j) \Phi^{(n)}_{2 k}(x^{(n)}_l) \right] 
\nonumber \\ 
& - & \frac{\Phi^{(m)}_{N-1}(x)}{s_{N-1}} \sum_{k=(N+1)/2}^{\infty} \frac{1}{r_k} \left[ 
s_{2 k} \Phi^{(n)}_{2 k + 1}(x^{(n)}_l) -   
s_{2 k + 1} \Phi^{(n)}_{2 k}(x^{(n)}_l) \right] 
\nonumber \\ 
& - & \frac{\Phi^{(n)}_{N-1}(y)}{s_{N-1}} \sum_{k=(N+1)/2}^{\infty} \frac{1}{r_k} \left[ 
\Phi^{(m)}_{2 k}(x^{(m)}_j) s_{2 k + 1} -   
\Phi^{(m)}_{2 k + 1}(x^{(m)}_j) s_{2 k} \right]. \nonumber \\ 
\end{eqnarray}
\par
Thus we have shown that the Pfaffian formulas are expressed 
in terms of the expansion coefficients  $\alpha_{jl}$ and 
$\beta_{jl}$. Let us explain how $\alpha_{jl}$ can be calculated. 
Suppose that ${\tilde R}_k(x) = x^k + \cdots$ are the monic skew 
orthogonal polynomials with the skew orthogonality relation    
\begin{eqnarray}
& & \langle {\tilde R}_{2 j},{\tilde R}_{2 l 
+ 1} \rangle^{(1)} = - \langle {\tilde R}_{2 
l + 1},{\tilde R}_{2 j} \rangle^{(1)} =  
{\tilde r}_j \delta_{jl},  \\  
& & \langle {\tilde R}_{2 j},
{\tilde R}_{2 l} \rangle^{(1)} = 0, \ \ \ 
\langle {\tilde R}_{2 j+1},{\tilde 
R}_{2 l + 1} \rangle^{(1)} = 0
\end{eqnarray}
and that ${\tilde R}_k(x)$ are expanded 
in terms of $C^{(1)}_j(x)$ as 
\begin{equation}
\label{rtkex}
{\tilde R}_k(x) 
= \sum_{j=0}^k {\tilde \alpha}_{kj} C^{(1)}_j(x), \ \ \ 
{\tilde \alpha}_{kk} = 1. 
\end{equation}
Since $R^{(1)}_k(x)$ should equal to ${\tilde R}_k(x)$ multiplied by a 
constant $c_k$, we can see that 
\begin{equation}
\alpha_{kj} = {\tilde \alpha}_{kj}  
\gamma^{(1)}_j \sqrt{h^{(1)}_j} c_k, \ \ \  
r_k = c_{2 k} c_{2 k + 1} {\tilde r}_k. 
\end{equation}
Noting $\alpha_{kk} = \gamma^{(M)}_k \sqrt{h^{(M)}_k}$, we find 
\begin{equation}
c_k =  \frac{\gamma^{(M)}_k 
\sqrt{h^{(M)}_k}}{\gamma^{(1)}_k \sqrt{h^{(1)}_k}}. 
\end{equation}
Thus the coefficients $\alpha_{kj}$ can be calculated 
from the expansion (\ref{rtkex}).
\par
Let us list the cases related to the classical orthogonal 
polynomials in which the expansions (\ref{rtkex}) are explicitly known. 
The inverse expansions are sometimes shown instead of (\ref{rtkex}). 
The inversions of the expansions are easy in all of the listed 
cases. For simplicity an abbreviated notations $C_n(x)$ and 
$h_n$ are used for $C^{(1)}_n(x)$ and $h^{(1)}_n$, respectively. 
The sign function ${\rm sgn}(x)$ is defined as
\begin{equation}
{\rm sgn}(x) = \left\{ \begin{array}{ll} 
1, & x > 0, \\ 0, & x = 0, \\ -1, & x < 0.
\end{array} \right.
\end{equation}  
\par
\medskip
\noindent
{\bf (1) Hermite}\cite{MEHTA89,BN,NW}
\par
\medskip
\noindent
When the measure is given by
\begin{equation}
\int {\rm d}\mu_1(x) = \int_{-\infty}^{\infty} {\rm d}x \ {\rm e}^{-x^2},
\end{equation}
the corresponding monic orthogonal polynomials are
\begin{equation}
\label{mhermite}
C_n(x) = \frac{1}{2^n} H_n(x), \ \ \ h_n = \frac{\sqrt{\pi} n!}{2^n}.
\end{equation}
Here $H_n(x)$ are the Hermite polynomials
\begin{equation}
H_n(x) = (-1)^n {\rm e}^{x^2} \frac{{\rm d}^n}{{\rm d} x^n} 
{\rm e}^{-x^2}.  
\end{equation}
The monic skew orthogonal polynomials ${\tilde R}_n(x)$ are 
given as follows.
\par
\medskip
\noindent
Case I: 
$$
F(x,y) = {\rm e}^{(x^2 + y^2)/2} {\rm sgn}(y-x),
$$
\begin{eqnarray}
{\tilde R}_{2 n}(x) & = & C_{2 n}(x), \nonumber \\ 
{\tilde R}_{2 n + 1}(x) & = & C_{2 n + 1}(x) - 
n C_{2 n - 1}(x), \nonumber \\ 
{\tilde r}_n & = & 2^{-2 n + 1} \sqrt{\pi} (2 n)!.   
\end{eqnarray}
\par
\medskip
\noindent
Case II:
$$
F(x,y) = 2 {\rm e}^{(x^2+y^2)/2} \frac{\partial}{\partial x}\delta(x - y),
$$
\begin{eqnarray}
C_{2 n}(x) & = & {\tilde R}_{2 n}(x) - n {\tilde R}_{2 n - 2}(x), \nonumber \\ 
C_{2 n + 1}(x) & = & {\tilde R}_{2 n + 1}, \nonumber \\ 
{\tilde r}_n & = & 2^{-2 n} \sqrt{\pi} (2 n + 1)!.   
\end{eqnarray} 
\par
\medskip
\noindent
{\bf (2) Laguerre}\cite{MEHTA89,NW,NF95,AFNM} 
\par
\medskip
\noindent
The measure is
\begin{equation}
\int {\rm d}\mu_1(x) = \int_0^{\infty} {\rm d}x \ x^a {\rm e}^{-x}.
\end{equation}
The corresponding monic orthogonal polynomials are
\begin{equation}
C_n(x) = (-1)^n n! L^{(a)}_n(x), \ \ \ h_n = n! (n + a)!,
\end{equation}
where $L^{(a)}_n(x)$ are the Laguerre polynomials
\begin{equation}
L^{(a)}_n(x) = \frac{x^{-a} {\rm e}^{x}}{n!} \frac{{\rm d}^n}{{\rm d} x^n} 
({\rm e}^{-x} x^{n + a}).  
\end{equation}
\par
\medskip
\noindent
Case I: 
$$
F(x,y) = 
x^{-(a+1)/2} y^{-(a+1)/2} {\rm e}^{(x + y)/2} 
{\rm sgn}(y-x),
$$
\begin{eqnarray}
{\tilde R}_{2 n}(x) & = & C_{2 n}(x), \nonumber \\ 
{\tilde R}_{2 n + 1}(x) & = & C_{2 n + 1}(x) - 
(2 n)(a + 2 n) C_{2 n - 1}(x), \nonumber \\ 
{\tilde r}_n & = & 4 (2 n)! (a + 2 n)!.
\end{eqnarray}
\par
\medskip
\noindent
Case II: 
$$
F(x,y) = 
2 x^{-(a-1)/2} y^{-(a-1)/2} {\rm e}^{(x+y)/2} 
\frac{\partial}{\partial x}\delta(x - y),
$$
\begin{eqnarray}
C_{2 n}(x) & = & {\tilde R}_{2 n}(x) - (2 n) (a + 2 n) 
{\tilde R}_{2 n - 2}(x), \nonumber \\ 
C_{2 n + 1}(x) & = & {\tilde R}_{2 n + 1}, \nonumber \\ 
{\tilde r}_n & = & (2 n + 1)! (a + 2 n + 1)!.   
\end{eqnarray} 
An interpolation between Case I and Case II was studied in \cite{FR}. 
\par
\medskip
\noindent
{\bf (3) Jacobi}\cite{MEHTA76,MEHTA89,NW,NF95,AFNM} 
\par
\medskip
\noindent
The measure is
\begin{equation}
\int {\rm d}\mu_1(x) = \int_{-1}^1 {\rm d}x \ (1 - x)^a (1 + x)^b.
\end{equation}
The corresponding monic orthogonal polynomials are
\begin{eqnarray}
C_n(x) & = & 2^n n! \frac{(a + b + n)!}{(a + b + 2 n)!} P^{(a,b)}_n(x), \nonumber \\ 
h_n & = & 2^{a + b + 2 n + 1} n! \frac{(a + n)! (b + n)!(a + b + n)!}{(a + b + 2 n)! 
(a + b + 2 n + 1)!},
\end{eqnarray}
where $P^{(a,b)}_n(x)$ are the Jacobi polynomials
\begin{equation}
P^{(a,b)}_n(x) = \frac{1}{(1 - x)^a (1 + x)^b} 
\frac{(-1)^n}{2^n n!} \frac{{\rm d}^n}{{\rm d} x^n} 
\left\{ (1 - x)^{a + n} (1 + x)^{b + n} \right\}.  
\end{equation}
\par
\medskip
\noindent
Case I: 
$$
F(x,y) = (1 - x)^{-(a+1)/2} (1 - y)^{-(a+1)/2} 
(1 + x)^{-(b+1)/2} (1 + y)^{-(b+1)/2} 
{\rm sgn}(y-x),
$$
\begin{eqnarray}
& & {\tilde R}_{2 n}(x) = C_{2 n}(x), \nonumber \\ 
& & {\tilde R}_{2 n + 1}(x) = C_{2 n + 1}(x) \nonumber \\ 
& - & 
\frac{8 n (a + 2 n) (b + 2 n) (a + b + 2 n)}{(a + b + 4 n - 
1)(a + b + 4 n)(a + b + 4 n + 1)(a + b + 4 n + 2)} C_{2 n - 1}(x), \nonumber \\ 
& & {\tilde r}_n = 2^{a + b + 4 n + 3} (2 n)! \frac{(a + 2 n)! (b + 2 n)! (a + b + 
2 n)!}{(a + b + 4 n)! (a + b + 4 n + 2)!}.
\end{eqnarray}
\par
\medskip
\noindent
Case II: 
$$
F(x,y) = 2 (1 - x)^{-(a-1)/2} (1 - y)^{-(a-1)/2} 
(1 + x)^{-(b-1)/2} (1 + y)^{-(b-1)/2} 
\frac{\partial}{\partial x}\delta(x - y),
$$
\begin{eqnarray}
& & C_{2 n}(x) = {\tilde R}_{2 n}(x) \nonumber \\ 
& - & \frac{8 n (a + 2 n) (b + 2 n) (a + b + 2 n)}{
(a + b + 4 n + 1)(a + b + 4 n)(a + b + 4 n -1)(a + b + 4 n-2)}
{\tilde R}_{2 n - 2}(x), \nonumber \\ 
& & C_{2 n + 1}(x) = {\tilde R}_{2 n + 1}, \nonumber \\ 
& & {\tilde r}_n = 2^{a + b + 4 n + 3} (2 n + 1)! \frac{(a + 2 n + 1)! (b + 2 n + 1)! 
(a + b + 2 n + 1)!}{(a + b + 4 n + 1)! (a + b + 4 n + 3)!}. \nonumber \\ 
\end{eqnarray}
An interpolation between Case I and Case II was studied in \cite{FR}. 
\par
\medskip
\noindent
{\bf (4) Symmetic Hahn}\cite{NF02} 
\par
\medskip
\noindent
The measure is
\begin{equation}
\int {\rm d}\mu_1(x) = \sum_{x = -\infty}^{\infty} 
\frac{1}{\displaystyle \left[ \left( \frac{L}{2} 
+ x \right)! \left( \frac{L}{2} - x \right)! \right]^2},
\end{equation}
where $x$ is an integer when $L$ is an even integer, while it 
is a half odd integer when $L$ is odd. The corresponding 
monic orthogonal polynomials are 
\begin{eqnarray}
C_n(x) & = & (-1)^n \frac{(L!)^2 (2 L - 2 n + 1)!}{\{ (L-n)!\}^2 
(2 L - n + 1)!} Q^{(-L-1,-L-1)}_n\left( x + 
\frac{L}{2} ; L \right), \nonumber \\ 
h_n & = & \frac{n! (2 L - 2 n  + 1)! (2 L - 2 n)!}{(2 L - n + 1 )! 
\left\{ (L - n)! \right\}^4},
\end{eqnarray}
where $Q^{(a,b)}_n(x;L)$ are the Hahn polynomials\cite{KS}
\begin{equation}
\label{hahn}
Q^{(a,b)}_n(x;L) = \sum_{k=0}^L \frac{(-n)_k (n + a + b + 1)_k (-x)_k}{
(a + 1)_k (-L)_k} \frac{1}{k!}, 
\end{equation}
\par
\medskip
\noindent
where $(a)_n = (a+n-1)!/(a-1)!$. If $F(x,y)$ is given by
$$
F(x,y) = 
\left( \frac{L}{2} + x \right)! \left(\frac{L}{2} - x \right)! 
\left( \frac{L}{2} + y \right)! \left(\frac{L}{2} - y \right)! 
\ {\rm sgn}(y-x),
$$
then
\begin{eqnarray}
& & {\tilde R}_{2 n}(x) = C_{2 n}(x), \nonumber \\ 
& & {\tilde R}_{2 n + 1}(x) = C_{2 n + 1}(x) - 
\frac{n (L - n + 1)(L - 2 n)(L- 2 n + 1)}{(2 L - 
4 n + 3)(2 L - 4 n + 1)} 
C_{2 n - 1}(x), \nonumber \\ 
& & {\tilde r}_n = 
\frac{(2 n)! (2 L - 4 n  + 1)! (2 L - 4 n)!}{2 (2 L - 2 n + 1 )! 
(L - 2 n - 1)! \left\{ (L - 2 n)! \right\}^3}.
\end{eqnarray}
\par
\medskip
\noindent
{\bf (5) Discrete Chebyshev} 
\par
\medskip
\noindent
The measure is
\begin{equation}
\int {\rm d}\mu_1(x) = \sum_{x = 0}^{L}, 
\end{equation}
where $x$ is an integer. The corresponding monic orthogonal 
polynomials are 
\begin{eqnarray}
C_n(x) & = &  (-1)^n \frac{L! (n!)^2}{(L-n)! (2 n)!} 
Q^{(0,0)}_n\left( x; L \right), \nonumber \\ 
h_n & = & \frac{(L + n + 1)! (n!)^4}{(L-n)! (2 n + 1)! (2 n)!},
\end{eqnarray}
where $Q^{(a,b)}_n(x;L)$ are the Hahn polynomials (\ref{hahn}). If $F(x,y)$ 
is given by
$$
F(x,y) = {\rm sgn}(y-x),
$$
then
\begin{eqnarray}
C_{2 n}(x) & = & {\tilde R}_{2 n}(x) - \frac{n (2 n - 1) (L - 2 n + 1) 
(L + 2 n + 1)}{2 (4 n - 1) (4 n + 1)} {\tilde R}_{2 n - 2}(x), \nonumber \\ 
C_{2 n + 1}(x) & = & {\tilde R}_{2 n + 1}(x), \nonumber \\ 
{\tilde r}_n & = & 2 \frac{(L + 2 n + 2)! (2 n)! \left\{ (2 n + 1)! 
\right\}^3}{(L - 2 n - 1)! (4 n + 2)! (4 n + 3)!}.     
\end{eqnarray} 
\par
\medskip
\noindent
{\bf (6) Discrete exponential}\cite{FNR}
\par
\medskip
\noindent
The measure is
\begin{equation}
\int {\rm d}\mu_1(x) = \sum_{x = 0}^{\infty} q^x, 
\end{equation}
where $x$ is an integer. The corresponding monic orthogonal 
polynomials are 
\begin{eqnarray}
C_n(x) & = & \frac{q^n n!}{(q-1)^n} 
M_n(x;1,q), \nonumber \\ 
h_n & = & \frac{q^n (n!)^2}{(1 - q)^{2 n + 1}},
\end{eqnarray}
where $M_n(x;c,q)$ are the Meixner polynomials\cite{KS} 
\begin{equation}
M_n(x;c,q) = {}_2F_1\left( \left. \begin{array}{c} -n,-x \\ 
c \end{array} \right| 1 - \frac{1}{q} \right).
\end{equation} 
If $F(x,y)$ 
is given by
$$
F(x,y) = q^{-(x+y)/2} \alpha^{|y-x|/2} {\rm sgn}(y-x),
$$
then
\begin{eqnarray}
{\tilde R}_{2n}(x) & = & C_{2 n}(x) 
+ \sum_{k=0}^{n-1} \frac{(2 n)!}{(2 k)!} 
\frac{q^{n-k}}{(1-q)^{2 n - 2 k}} C_{2 k}(x) \nonumber \\ 
& - & \frac{\sqrt{\alpha} - \sqrt{q}}{1 - \sqrt{\alpha q}} 
\sum_{k=0}^{n-1} \frac{(2 n)!}{(2 k+1)!} 
\frac{q^{n-k-(1/2)}}{(1-q)^{2 n - 2 k-1}} C_{2 k+1}(x), \nonumber \\  
{\tilde R}_{2 n + 1}(x) & = & 
C_{2 n + 1}(x) - \frac{\sqrt{\alpha}-\sqrt{q}}{1 - 
\sqrt{\alpha q}} \frac{\sqrt{q}}{1-q} (2 n + 1) C_{2 n}(x), \nonumber \\ 
{\tilde r}_n & = & \frac{(2 n)! (2 n + 1)! 
\sqrt{\alpha} q^{2 n + (1/2)}}{(1 - q)^{
4 n + 1} (1 - \sqrt{\alpha q})^2}.
\end{eqnarray}

\section{Relevance to Random Matrix Theory}
\setcounter{equation}{0}
\renewcommand{\theequation}{5.\arabic{equation}}

In this section we explain how the Pfaffian formulas are 
applied in random matrix theory and other applications. 
One of the most popular examples is the Gaussian orthogonal 
ensemble (GOE)\cite{MEHTA04}. The GOE is an ensemble of $N 
\times N$ real symmetric matrices $X$ with a Gaussian probability 
distribution function (p.d.f.)
\begin{equation}
\label{gaussian}
P(X) {\rm d}X \propto {\rm exp}\left\{ - \frac{1}{2} {\rm Tr} X^2 \right\} {\rm d}X,
\end{equation}
where
\begin{equation}
{\rm d}X = \prod_{j=1}^N {\rm d}X_{jj} \prod_{j<l}^N {\rm d}X_{jl}. 
\end{equation}
\par
Let us change the variables from the matrix elements of $X$ to 
the eigenvalues $x_1,x_2,\cdots,x_N$ and eigenvector parameters. 
Integrating out the eigenvector parameters, we find the p.d.f. 
of the eigenvalues
\begin{equation}
p(x_1,x_2,\cdots,x_N) \prod_{j=1}^N {\rm d}x_j \nonumber \\ 
\propto \prod_{j=1}^N {\rm e}^{ - \frac{1}{2} (x_j)^2 }
\prod_{j<l}^N |x_j - x_l| \prod_{j=1}^N {\rm d}x_j.
\end{equation}
Therefore, we obtain
\begin{equation}
p(x_1,x_2,\cdots,x_N) \propto \prod_{j=1}^N {\rm e}^{ - \frac{1}{2} (x_j)^2 } 
\prod_{j>l}^N (x_j - x_l) {\rm Pf}[{\rm sgn}(x_l - x_j)]_{j,l=1,2,\cdots,N} 
\end{equation}
for even $N$ and 
\begin{eqnarray}
& & p(x_1,x_2,\cdots,x_N) \nonumber \\ 
& \propto & \prod_{j=1}^N {\rm e}^{ - \frac{1}{2} (x_j)^2 } 
\prod_{j>l}^N (x_j - x_l) {\rm Pf}\left[ \begin{array}{cc} 
[{\rm sgn}(x_l - x_j)]_{j,l=1,2,\cdots,N} & [1]_{j=1,2,\cdots,N} 
\\ - \left[1 \right]_{l=1,2,\cdots,N} & 0 \end{array} \right] \nonumber \\ 
\end{eqnarray}
for odd $N$. These are clearly special cases of the general formulas (\ref{peven}) and 
(\ref{podd}) with $M = 1$. The measure and $F(x,y)$ are given in Case I of 
{\bf (1) Hermite} in previous section. Therefore the correlation functions 
of the eigenvalues 
\begin{equation}
\rho(x_1,\cdots,x_k) = 
\frac{N!}{(N-k)!} \int \prod_{j=k + 1}^N {\rm d}x_j \ p(x_1,x_2,\cdots,x_N) 
\end{equation}
can be evaluated in Pfaffian forms.
\par
The ensemble of self-dual real quaternion random matrices with a Gaussian p.d.f. 
is called the Gaussian symplectic ensemble(GSE). The eigenvalue correlations of the 
GSE can similarly be analyzed using the formulas given in Case II of {\bf (1) Hermite} 
in previous section. It is also possible to express the correlation functions 
for a similar ensemble of antisymmetric hermitian matrices in Pfaffian forms: 
the derivation of the corresponding skew orthogonal polynomials is illustrated 
in \cite{NF98}. 
\par
The eigenvalue distributions of the matrices of the form $A^{\dagger} A$, 
where $A$ are $M \times N$ rectangular random matrices with 
Gaussian-distributed elements, are often important 
in applications. Here $A^{\dagger}$ is the hermitian conjugate of $A$. 
When $A$ are real matrices, we can employ the formulas in Case I of {\bf (2) 
Laguerre} to write the correlation functions in Pfaffian forms. When $A$ 
are real quaternion matrices, Pfaffian expression for the correlation 
functions are deduced from the formulas in Case II of {\bf (2) Laguerre}. 
\par
Hereafter we assume that $N$ is even for simplicity. The Pfaffian formulas 
are also useful in analyzing the dynamical correlation among the eigenvalues of the 
matrix Brownian motion model. The matrix Brownian motion model was also introduced 
by Dyson\cite{DYSON62}. He considered an $N \times N$ hermitian 
random matrix $X$ with the p.d.f.
\begin{equation}
P({X}^{(0)};{X};\tau) {\rm d}X \propto
{\rm exp}\left[- \frac{
{\rm Tr} \left\{ (
X  - {\rm e}^{-\tau} X^{(0)})^2 \right\}}{1 
- {\rm e}^{-2 \tau}} \right] {\rm d}X
\end{equation}
depending on the parameter $\tau$. Here $X^{(0)}$ 
is another matrix whose symmetry belongs to a subclass of the 
symmetry of $X$. Note that $X$ is equated with $X^{(0)}$ 
at $\tau = 0$. The measure is chosen to be
\begin{equation}
{\rm d}X = \prod_{j=1}^N {\rm d}X_{jj} 
\prod_{j<l}^N {\rm d}{\rm Re}X_{jl} \ {\rm d}{\rm Im}X_{jl}. 
\end{equation} 
\par
Dyson showed that the eigenvalue p.d.f. $p$ of $X$ satisfies 
the Fokker-Planck equation
\begin{equation}
\frac{\partial p}{\partial \tau} = {\cal L}p, \ \ \ 
{\cal L} = \frac{1}{2} \sum_{j=1}^N \frac{\partial}{\partial x_j}
{\rm e}^{-2 w} \frac{\partial}{\partial x_j} {\rm e}^{2 w},
\end{equation}
where $x_1,x_2,\cdots,x_N$ are the eigenvalues of $X$ and
\begin{equation}
w = \frac{1}{2} \sum_{j=1}^N x_j^2 - \sum_{j<l}^N
\log |x_j - x_l |.
\end{equation}
In order to solve the Fokker-Planck equation, we need 
to specify the initial condition. For example, let us suppose 
that $X^{(0)}$ is a GOE random matrix. Then the Fokker-Planck equation 
can exactly be solved and the p.d.f. $p$ of the eigenvalues 
$x^{(m)}_1,x^{(m)}_2, \cdots,x^{(m)}_N$ at $\tau = \tau_m$, 
$m = 1,2,\cdots,M$, can be explicitly derived. Using the 
notation ${\bf x}^{(m)} = (x^{(m)}_1,x^{(m)}_2, \cdots,x^{(m)}_N)$, 
we find
\begin{eqnarray}
& & p({\bf x}^{(1)},{\bf x}^{(2)},
\cdots,{\bf x}^{(M)}) \prod_{m=1}^M \prod_{j=1}^N {\rm d}x^{(m)}_j 
\nonumber \\ 
& \propto & \prod_{j=1}^N {\rm e}^{-(x^{(M)}_j)^2/2} 
\prod_{j>l}^N (x^{(M)}_j - x^{(M)}_l) 
{\rm Pf}[F(x^{(1)}_j,x^{(1)}_l;\tau_1)]_{j,l = 1,2,\cdots,N} \nonumber \\ 
& \times & 
\prod_{m=1}^{M-1} {\rm det}[g(x^{(m+1)}_j,x^{(m)}_l;\tau_{m+1}-\tau_m)]_{j,l = 
1,2,\cdots,N} \prod_{m=1}^M \prod_{j=1}^N {\rm d}x^{(m)}_j. 
\nonumber \\ 
\end{eqnarray}
Here
\begin{equation}
F(x,y; \tau) = \int_{-\infty}^{\infty} {\rm d}z^{\prime} 
\int_{-\infty}^{z^{\prime}} 
{\rm d}z \{ g(x,z;\tau) g(y,z^{\prime};\tau^{\prime}) - 
g(y,z;\tau^{\prime}) g(x,z^{\prime};\tau) \}
\end{equation}
with
\begin{equation}
g(x,y;\tau) = {\rm e}^{-(x^2+y^2)/2} \sum_{j=0}^\infty
\frac{C_j(x)C_j(y)}{h_j} {\rm e}^{-\gamma_j\tau},
\end{equation}
where $C_n(x)$ and $h_n$ are defined in (\ref{mhermite}) 
and $\gamma_j = j + (1/2)$. 
The dynamical correlation functions 
\begin{eqnarray}
\label{dcf}
& & \rho(x^{(1)}_1,\cdots,x^{(1)}_{k_1};x^{(2)}_1,\cdots,x^{(2)}_{k_2}; 
\cdots ;x^{(M)}_1,\cdots,x^{(M)}_{k_M}) = \frac{(N!)^M}{\prod_{m=1}^M (N - k_m)!} \nonumber \\ 
& \times & \int \prod_{j=k_1 + 1}^N {\rm d}x^{(1)}_j
\int \prod_{j=k_2 + 1}^N {\rm d}x^{(2)}_j \cdots
\int \prod_{j=k_M + 1}^N {\rm d}x^{(M)}_j p({\bf x}^{(1)},{\bf x}^{(2)},\cdots,{\bf x}^{(M)}) 
\nonumber \\ 
\end{eqnarray}
are then calculated in Pfaffian forms. This Pfaffian formula is 
known to be useful in analyzing the asymptotic behavior of the 
dynamical correlation functions\cite{MEHTA04,FNH,KT}.  
\par
The Pfaffian formulas can also be applied to multi-matrix models in which 
one $N \times N$ real symmetric matrix $X^{(1)}$ is combined to $N \times N$ 
hermitian matrices $X^{(2)},X^{(3)},\cdots, X^{(M)}$\cite{NAGAO01}. 
The p.d.f. is given by 
\begin{eqnarray}
& & P(X^{(1)},X^{(2)},\cdots,X^{(M)}) ({\rm d}X) \nonumber \\ 
& \propto & {\rm exp}\left[ - {\rm Tr}\left\{ 
\frac{1}{2} V_1(X^{(1)}) + \sum_{m=2}^{M-1}V_m(X^{(m)}) 
+ \frac{1}{2} V_M(X^{(M)}) \right\} \right] 
\nonumber \\ 
& \times & 
{\rm exp}\left[ {\rm Tr}\left\{ c_1 X^{(1)} X^{(2)} 
+ c_2 X^{(2)} X^{(3)} + \cdots + c_{M-1} X^{(M-1)} X^{(M)} 
\right\} \right] ({\rm d}X) \nonumber \\ 
\end{eqnarray}
with the measure
\begin{eqnarray}
({\rm d}X) & = &  
\prod_{j=1}^N {\rm d}X^{(1)}_{jj} \prod_{j<l}^N {\rm d}X^{(1)}_{jl} 
\nonumber \\ & \times &  
\prod_{m=2}^M \left\{ \prod_{j=1}^N {\rm d}X^{(m)}_{jj} 
\prod_{j<l}^N {\rm d}{\rm Re}X^{(m)}_{jl} {\rm d}{\rm Im}X^{(m)}_{jl} 
\right\}. 
\end{eqnarray} 
\par
As before we like to change the variables and integrate out the eigenvector parameters. 
For that purpose, Harish-Chandra's integral formula can be utilized: for $N \times N$ 
diagonal matrices $A$ and $B$ with diagonal elements $a_1,a_2,\cdots,a_N$ and 
$b_1,b_2,\cdots,b_N$, respectively, the integral over the group of $N \times N$ 
unitary matrices $U$ can be evaluated as\cite{HC}
\begin{eqnarray}
& & \int {\rm exp}\left\{ - \frac{1}{t} {\rm Tr} (A - U B U^{\dagger})^2 \right\} 
{\rm d}U 
\nonumber \\ & \propto & \frac{1}{\prod_{j<l}^N (a_j - a_l)(b_j - b_l)} {\rm det}\left[
{\rm exp}\left\{ - \frac{1}{t} (a_j - b_l)^2 \right\} \right]_{j,l=1,2,\cdots,N}.
\end{eqnarray}
\par
Let us write the eigenvalues of $X^{(m)}$ as 
$x^{(m)}_1,x^{(m)}_2,\cdots,x^{(m)}_N$ 
and define the vectors ${\bf x}^{(m)} = 
(x^{(m)}_1,x^{(m)}_2,\cdots,x^{(m)}_N)$. Then, after 
integrating out the eigenvector parameters, we obtain the p.d.f. of the eigenvalues
\begin{eqnarray}
\label{pmm}
& & p({\bf x}^{(1)},{\bf x}^{(2)},\cdots,{\bf x}^{(M)}) 
\prod_{m=1}^M \prod_{j=1}^N {\rm d}x^{(m)}_j 
\nonumber \\ 
& \propto & \prod_{j>l}^N (x^{(M)}_j - x^{(M)}_l) \ 
{\rm Pf}[{\rm sgn}(x^{(1)}_l-x^{(1)}_j)]_{j,l=1,2,\cdots,N} 
\nonumber \\ & \times &   
\prod_{m=1}^{M-1} 
\det[g^{(m)}(x^{(m+1)}_j,x^{(m)}_l)]_{j,l = 1,2,\cdots,N} \prod_{m=1}^M \prod_{j=1}^N 
{\rm d}x^{(m)}_j 
\end{eqnarray}
with
\begin{equation}
g^{(m)}(x,y) = {\rm exp}\left\{ -\frac{1}{2} V_{m+1}(x) - \frac{1}{2} V_m(y) + c_m x y 
\right\}.
\end{equation}
The correlation functions for the multi-matrix models are
defined by the formula (\ref{dcf}) with the p.d.f. $p$ replaced by (\ref{pmm}). 
Comparison of the above formula with (\ref{peven}) reveals that the correlation 
functions are written in Pfaffian forms.
\par
Recently, through the discretization of random matrix ensembles, it turned out 
that the dynamical correlation functions for stochastic motions of 
interacting many particles were sometimes written in Pfaffian forms. 
One of the simplest examples is given by the vicious walkers\cite{NF02,NKT} 
on the lattice ${\bf Z} = \{\cdots,-2,-1,0,1,2,\cdots\}$. Let us consider 
$N$ simple symmetric random walkers who start from the positions 
$2 X_1 < 2 X_2 < \cdots < 2 X_N, \ X_j \in {\bf Z}$, at time $k = 0$. 
The positions $S^{(k)}_j$ of the $j$-th walkers at time $k \geq 0$ are restricted 
by the nonintersecting condition 
\begin{equation}
S^{(k)}_1 < S^{(k)}_2 < \cdots < S^{(k)}_N, \ \ \ 1 \leq k \leq K.
\end{equation}
Let $V(X_1,\cdots,X_N;Y_1,\cdots,Y_N)$ be the probability 
that the $N$ walkers come to 
the positions $2 Y_1 < 2 Y_2 < \cdots < 2 Y_N, \ Y_j 
\in {\bf Z}$, at an even integer time $K$. Then the 
nonintersecting condition implies a determinant formula 
\begin{equation}
\label{vxy}
V(X_1,\cdots,X_N;Y_1,\cdots,Y_N)=\frac{1}{2^{KN}} \det\left[\left( 
\begin{array}{c} K \\ 
K/2 + X_l - Y_j \end{array} \right) \right]_{j,l=1,2,\cdots,N}. 
\end{equation}
Hereafter we employ the scaled variables $t = K/L$, $x_j = 
2 X_j/\sqrt{L}$ and $y_j = 2 Y_j \sqrt{L}$, with which a 
simplification occurs in the scaling limit $L \rightarrow \infty$.
\par
We consider the nonintersecting motion in the scaled time interval 
$0 < t \leq T$ and suppose that the scaled positions of all 
the walkers are initially at the origin. Let us denote the scaled 
positions of the $j$-th walker at a scaled time $t_m$ by $x^{(m)}_j$ 
and define ${\bf x}^{(m)} = (x^{(m)}_1,x^{(m)}_2,\cdots,x^{(m)}_N)$. 
Then, from the determinant formula (\ref{vxy}), the p.d.f. of the 
walkers is evaluated in the limit $L \rightarrow \infty$ as 
\begin{eqnarray}
\label{pvicious} 
& & p({\bf x}^{(1)},{\bf x}^{(2)},\cdots,{\bf x}^{(M)}) 
\prod_{m=1}^M \prod_{j=1}^N {\rm d}x^{(m)}_j 
\nonumber \\ & \propto & 
\int \prod_{j=1}^N {\rm d}x^{(M+1)}_j 
\prod_{j>l}^N (x^{(1)}_j - x^{(1)}_l)  
{\rm Pf}[{\rm sgn}(x^{(M+1)}_l - x^{(M+1)}_j)]_{j,l=1,2,\cdots,N} 
\nonumber \\ & \times & 
\prod_{m=1}^M \det[g^{(m)}(x^{(m)}_j,x^{(m+1)}_l)]_{j,l=1,\cdots,N} 
\prod_{m=1}^M \prod_{j=1}^N {\rm d}x^{(m)}_j, 
\end{eqnarray}
where 
\begin{eqnarray}
g^{(1)}(x,y) & = & \frac{{\rm e}^{-x^2/(2 t_1)} {\rm e}^{-(x-y)^2/(2(t_2-t_1))}}
{\sqrt{2 \pi t_1} \sqrt{2 \pi(t_2-t_1)}}, \nonumber \\ 
g^{(m)}(x,y) & = & \frac{{\rm e}^{-(x-y)^2/(2(t_{m+1}-t_{m}))}}
{\sqrt{2 \pi (t_{m+1}-t_{m})}}, \ \ 2 \leq  m \leq M 
\end{eqnarray}
with $t_{M+1} = T$. The dynamical correlation functions can be defined 
by the formula (\ref{dcf}) with the p.d.f. $p$ replaced by (\ref{pvicious}). 
Then we can apply the Pfaffian formulas in \S 2. The asymptotic behavior 
of the correlation functions can be analyzed by using the Pfaffian 
formulas\cite{NKT}. 
\par
In this article we focused on the applications of the Pfaffian 
to hermitian random matrices and related stochastic models. 
It should be remarked that the Pfaffian is also useful in 
evaluating the eigenvalue correlations of unitary\cite{NF03}, 
non-self-dual real quaternion\cite{MEHTA04} and 
asymmetric real\cite{FN} random matrices.

\end{document}